\documentclass[aps,pra,showpacs,notitlepage]{revtex4-1}

\usepackage{graphicx}

\begin{document}

\title{How quantum paradoxes originate from the non-classical statistics of physical properties related to each other by half-periodic transformations}

\author{Holger F. Hofmann}
\email{hofmann@hiroshima-u.ac.jp}
\affiliation{
Graduate School of Advanced Sciences of Matter, Hiroshima University,
Kagamiyama 1-3-1, Higashi Hiroshima 739-8530, Japan}
\affiliation{JST, CREST, Sanbancho 5, Chiyoda-ku, Tokyo 102-0075, Japan
}

\begin{abstract}
Quantum paradoxes show that quantum statistics can exceed the limits of positive joint probabilities for physical properties that cannot be measured jointly. It is therefore impossible to describe the relations between the different physical properties of a quantum system by assigning joint realities to their observable values. Instead, recent experimental results obtained by weak measurements suggest that non-classical correlations could be expressed by complex valued quasi-probabilities,  where the phases of the complex probabilities express the action of transformations between the non-commuting properties (H. F. Hofmann, New J. Phys. {\bf 13}, 103009 (2011)).  In these relations, negative probabilities necessarily emerge whenever the physical properties involved are related to each other by half-periodic transformations, since such transformations are characterized by action phases of $\pi$ in their complex probabilities. It is therefore possible to trace the failure of realist assumptions back to a fundamental and universally valid relation between statistics and dynamics that associates half-periodic transformations with negative probabilities.
\end{abstract}


\maketitle

\section{Introduction}

Among all theories of physics, only quantum mechanics fails to describe the state of a physical object in terms of its experimentally observable properties. This failure is often explained by invoking ``uncertainties'', which suggests that there could be a more complete description where quantum states can be interpreted as joint probabilities of the unknown properties. However, such explanations are not consistent with the actual Hilbert space formalism, as highlighted by a number of quantum paradoxes which demonstrate that the quantum state cannot be interpreted as a positive valued probability distribution of the potential measurement outcomes \cite{Bell,KS,LGI,box,Har92,Che}. Since quantum paradoxes are predicted by the standard formalism, one might expect that there should be a well explained reason for the failure of simple realist models. However, most discussions of quantum paradoxes treat the formalism as a black box that does not relate to any known aspects of physics - almost as if the quantum states were miraculous descriptions of disembodied knowledge, rather than actual physical conditions produced in a laboratory. 

Recently, there have been a number of experimental breakthroughs in the attempts to lift the veil behind which the physics of quantum paradoxes is hiding. By using weak measurements, researchers in a number of laboratories have demonstrated that it is possible to obtain non-positive joint probabilities from the experimental data \cite{Res04,Jor06,Tol07,Wil08,Lun09,Yok09,Gog11,Suz12,Den14}. Oddly, these experiments have not led to a more thorough analysis of the physics described by those negative probability results, even though it should be the main purpose of quantum theory to explain experimental results, not just to simulate them by otherwise unexplained mathematical formulas. The problem might be that theorists have been distracted by the formal aspects of information and statistics, while forgetting entirely that, realistic physical situations are implemented through processes that are described by interactions and not by disembodied states of knowledge. In the context of weak measurements, this resulted in the misconception that weak values are ``strange'', merely because they can exceed the limits set by the eigenvalue spectrum \cite{Aha88,Hos10}. However, the outcomes of weak measurements are a natural consequence of physical interactions, and their strangeness is merely an indication that we have misunderstood the physics described by the standard formalism. In fact, the ``strangeness'' of weak values originates from the same misguided tendency to explain quantum statistics by assigning artificial realities that is causing the confusion in the discussion of quantum paradoxes. It would be good to realize just how artificial the present discussion of ``models'' in quantum mechanics has become. We should not neglect the actual physics described by the operator algebra while trying to explain its results by adding additional layers to the theory without any physical justification. Instead, we should try to understand the actual physics described by the quantum formalism, focusing on the fundamental patterns that produce all of these ``strange'' results. 

Since both weak measurements and quantum paradoxes are concerned with statistical relations between physical properties that cannot be measured jointly, the correct question to ask is this: what is the relation between the physical properties of an object that prevents any joint measurement? Ironically, part of the answer is well known from standard quantum mechanics. Properties cannot be measured jointly if they are represented by non-commuting operators with no joint eigenstates. In terms of physics, non-commutativity means that the dynamics generated by one property will transform the other physical property and vice versa. Since this dynamical relation between physical properties means that a measurement interaction sensitive to one will transform the other, there is no empirical foundation for the assumption of a joint reality of these physical properties. It is therefore quite reasonable that the fundamental relation between these properties is given by complex-valued probabilities that express both the impossibility of a joint reality and the dynamical laws of physics that replace the functional dependence of joint realities \cite{Hof11,Hof12,Hof14a}. 

It may be important to emphasize just how different this approach is from the present discussions of quantum paradoxes as limits to alternative theories. As I will discuss in more detail below, the conventional formalism creates the false impression of interpretational freedom by artificially separating the state of a system from its properties. This separation makes it easy to ignore experimental evidence, e.g. by failing to acknowledge that weak and strong measurements originate from the same class of interactions with only a single well-defined property. Thus the key to a proper explanation of quantum paradoxes is found in a more consistent description of the specific physical properties involved, independent of states or measurements. I would like to propose that complex probabilities provide such a consistent description by expressing the universal and context-independent relation between physical properties in a way that can be applied equally to state preparation, propagation, and measurement. 

In the following, I will show that the negative probabilities observed in weak measurement experiments can be explained by the structure of transformations in the standard quantum formalism. Specifically, the complex probabilities obtained in weak measurements actually express the action of transformations between the three physical properties defined by initial state, final state, and weak measurement, establishing a fundamental relation between the mathematical form of reversible transformations and the mathematical form of non-classical correlations in the operator formalism \cite{Hof11,Hof12,Hof14a}. Negative probabilities emerge whenever the action phases that describe the transformations between the physical properties are larger than $\pi/2$. Maximal negative probabilities are observed whenever the action phases of the complex probabilities are $\pi$, which is typical for half-periodic transformations. Quantum paradoxes are therefore naturally observed in the correlations between physical properties that are related to each other by half-periodic transformations, such as spin flips and swap operations. 

The explanation of non-positive statistics by action phases can provide a much clearer picture of the conditions under which the naive assumption of a joint reality fails. In actual experiments, a joint observation of non-commuting properties is prevented by the measurement back-action, which corresponds to a randomization of the action phases in the complex probabilities. Once these random errors are accounted for, the experimental data obtained from sequential measurements confirms the negative probabilities associated with action phases of $\pi/2$ \cite{Suz12}. Moreover, the expression of quantum statistics by complex action phase probabilities also explains why the operator algebra describes the dynamics of expectation values as the imaginary part of an operator-valued correlation \cite{Bus13}. Thus the complex probabilities merely flesh out the microscopic details that are hidden by a formalism that distinguishes quantum states from the physical properties that they ought to describe. 

In the discussion below, I will first show that non-positive probabilities emerge naturally from the conventional operator algebra that is commonly used to describe the statistics of quantum systems. I will then identify the fundamental relation between the dynamics of transformations and the joint probabilities of non-commuting properties. With this relation, it is possible to explain the inequality violations observed in specific quantum paradoxes, as will be discussed in detail in the main part of the paper. Finally, I consider the implication of the results for our fundamental understanding of quantum physics, pointing out that all paradoxes can be resolved by recognizing that there is no reality without the dynamics induced by measurement interactions. The explanation of quantum paradoxes by half-periodic transformations between the physical properties may thus close a significant gap in our present understanding of quantum physics.

\section{The structure of quantum paradoxes}

Quantum paradoxes concern the relation between physical properties that cannot be measured jointly. Standard quantum mechanics limits actual physical situations to the preparation of a state $a$ with a well defined physical property $A$, and the subsequent measurement of an outcome $m$ for a different physical property $M$, where the relation between $a$ and $m$ is described by the conditional probability $P(m|a)$ of the measurement outcomes $m$. Alternatively, it is possible to measure a third property $B$ with outcomes of $b$ distributed according to the conditional probability $P(b|a)$. However, the standard formalism indicates that there exists no joint measurement of $m$ and $b$ if the properties $M$ and $B$ do not have any common eigenstates. It is therefore impossible to measure a joint probability of the form $P(m,b|a)$ to directly characterize the relation between the three physical properties $A$, $M$ and $B$. 

All quantum paradoxes are based on the assumption that the measurement statistics of separate measurements should be consistent with a positive joint probability $P(m,b|a)$, even if it is impossible to obtain any joint measurement outcomes $(m,b)$. It is then shown that the combination of measurement statistics observed in separate measurements of the marginal probabilities defined by $P(m,b|a)$ exceeds the limits obtained by assuming that all $P(m,b|a)$ must be positive real numbers. In general, quantum paradoxes therefore demonstrate that the statistics of quantum states cannot be reproduced by positive valued joint probabilities of pairs of measurement outcomes $(m,b)$.

To some extend, this is not a surprising result. If it was so simple to replace the rather unintuitive rules of quantum mechanics with a conventional joint probability, it would seem to be preferable to explain quantum physics in terms of the fundamental realities $(m,b)$, and not by some mysterious ``superpositions'' of mutually exclusive alternatives. However, the quantum formalism is not a black box, and there are well defined rules that determine the circumstances under which quantum statistics will exceed the limits imposed by positivity. It is therefore possible to explain quantum paradoxes as a natural consequence of the fundamental physics described by quantum theory. To do so, it is necessary to identify the universally valid relations between non-commuting properties that are encoded in the abstract mathematical formalism of the theory. Instead of separating formalism and intuition, we should thus develop a practical understanding of the fundamental physics described by quantum theory. 

In the following, I will show that quantum mechanics does provide a universally valid description of the relations between physical properties in the form of complex probabilities that incorporate the dynamical structure of physics \cite{Hof11,Hof12,Hof14a,Hof14b}. A quantum state can then be identified with a dynamical randomization along the trajectory generated by the associated physical property $a$. The joint probabilities $P(m,b|a)$ then arise from the universal relation between the three physical properties $A$, $M$ and $B$, which replaces and supersedes the classical relations by which $m$ would be determined by the intersection of the trajectory of $a$ and the trajectory of $b$ in phase space \cite{Hof12}. 

\section{Operator algebra and the physical meaning of complex probabilities}
\label{sec:algebra}

How does the quantum formalism describe the relation between non-commuting properties? In the conventional formulation, the outcomes of each measurement can be represented by projection operators, e.g. $\mid a \rangle \langle a \mid$ for the outcome $a$ of the property $A$, etc. The conditional probabilities $P(m|a)$ are then obtained from the product trace of the projection operators,
\begin{equation}
\label{eq:Born1}
P(m|a) = \mbox{Tr}\left( \mid m \rangle \langle m \mid a \rangle \langle a \mid \right) = P(a|m).
\end{equation}
In this relation, state preparation and measurement are expressed by the same operators, so that it is possible to exchange the roles of the two to obtain $P(a|m)=P(m|a)$. State preparation therefore corresponds to the selection of a specific physical property that characterizes the initial conditions in the experimental setup. 

Eq.(\ref{eq:Born1}) also shows that the relation between the two properties $A$ and $M$ includes an element of randomness. In general, each state $a$ can produce every possible outcome $m$, so there is no fundamental relation that determines $m$ as a function of $a$. Effectively, $a$ and $m$ are independent of each other, where $P(a|m)=P(m|a)$ evaluates how often $m$ and $a$ coincide. It would be tempting to identify this coincidence with a phase space point, but such a point would also need to explain all other experimental distributions, e.g. for a property $B$ that does not share any eigenstates with either $A$ or $B$,
\begin{equation}
\label{eq:Born2}
P(b|a) = \mbox{Tr}\left( \mid b \rangle \langle b \mid a \rangle \langle a \mid \right) = P(a|b).
\end{equation}
Quantum mechanics thus provides separate descriptions of the relations between $m$ and $a$ and between $b$ and $a$. The question is how quantum mechanics describes the relations between the three physical properties $b$, $m$ and $a$. Experimentally, a direct observation of this relation would require a joint measurement of $m$ and $b$ following a preparation of $a$, so it is tempting to assume that the question cannot be decided experimentally. However, experiments do not need to be fully resolved, and this possibility is used to investigate the relation between non-commuting properties in weak measurements \cite{Hof12}. The relation so obtained is mathematically quite simple and corresponds to the natural extension of Eqs.(\ref{eq:Born1},\ref{eq:Born2}) to products of the three projectors \cite{Hof14b},
\begin{equation}
\label{eq:jointprob}
P(m,b|a) = \mbox{Tr}\left( \mid b \rangle \langle b \mid m \rangle \langle m \mid a \rangle \langle a \mid \right),
\end{equation}
where the operator ordering corresponds to a weak measurement of $m$ followed by a precise measurement of $b$.
Interestingly, the similarity of this joint probability with classical phase space distributions was already noticed in the early days of quantum mechanics \cite{McCoy32,Dir45}. In particular, Dirac was the first to show that any operator (and hence any physical property) can be expressed as a function of two mutually overlapping basis sets $a$ and $b$ by using the weak values of the operator for these states \cite{Hof12,Dir45,Joh07}, and this relation has recently been used to perform quantum state tomography by weak measurements \cite{Lun12,Wu13,Sal13,Bam14}. 

The possibility of quantum tomography illustrates that the initial property $a$ can be completely defined in terms of the combinations of measurement outcomes $m$ and $b$. It is therefore possible to argue that the relation between $a$, $m$, and $b$ expressed by Eq.(\ref{eq:jointprob}) is both deterministic and universal \cite{Hof12,Hof14a}. Specifically, the approximate classical limit of Eq.(\ref{eq:jointprob}) would correspond to an ergodic distribution of phase space points $(m,b)$ along the trajectory generated by a Hamiltonian $A$ at a constant value of $a$. Eq.(\ref{eq:jointprob}) therefore expresses the fundamental relation between the physical properties $a$, $m$, and $b$ that replaces and corrects the classical trajectory, where $a$ corresponds to a set of points $(m,b)$. Significantly, this relation is universally valid and is neither random nor dependent on experimental circumstances. A negative value of $P(m,b|a)$ therefore describes a necessary relation between the measurement statistics of $P(m|a)$ and $P(b|a)$ that is determined only by the dynamical structure that replaces the classical phase space in the quantum limit. 

Since the projection operators do not commute with each other, the product trace of three operators that describes the joint probabilities in Eq.(\ref{eq:jointprob}) generally results in a complex number. Clearly, these complex numbers cannot be identified with relative frequencies of joint measurement outcomes, highlighting the fact that these probabilities describe a relation between physical properties that cannot have a joint reality. Experimentally, any attempt to perform simultaneous measurements of $m$ and $b$ under the initial condition $a$ will result in statistical uncertainties determined by the interaction dynamics of the measurement, and the consistency of the quantum formalism guarantees that the negative and imaginary contributions will always be ``covered up'' by measurement uncertainties. The situation is particularly clear in sequential measurements \cite{Suz12,Hof14c}: the intermediate measurement is characterized by a trade-off between measurement errors due to limited measurement resolution and back-action errors due to changes in the observable measured in the final measurement. In general, measurements require some kind of interaction dynamics, and the rules of quantum mechanics do not permit a clear separation between the measurement outcome and the dynamics generated by the observable in the system. The latter observation actually provides complex probabilities with a clear physical meaning: the complex phase represents the effects of transformation dynamics generated by one property on the probabilities of measurement outcomes for the other \cite{Hof11,Hof12,Hof14a}. 

It is now possible to proceed to the central insight of the present paper. Since the fundamental relation between physical properties in quantum mechanics is fundamentally defined by the complex joint probabilities $P(m,b|a)$, quantum paradoxes can be explained by identifying the conditions under which this relation results in negative joint probabilities. We can then expect to find that this kind of relation between physical properties appears in all known quantum paradoxes. In its original form as product trace of three projectors, the physics described by $P(m,b|a)$ is somewhat unclear. However, the same expression obtains a well-defined physical meaning if we consider how the dynamics generated by $m$ changes the probability of finding $b$ in $a$. In the standard formalism, the effects of a unitary transformation characterized by the action $S(m)$ assigned to the generator outcomes $m$ is described by
\begin{eqnarray}
\label{eq:unitary}
P(b|U(a)) &=&  |\langle b \mid \hat{U} \mid a \rangle|^2
\nonumber \\
&=& \left| 
\sum_m \langle b \mid m \rangle \langle m \mid a \rangle \exp\left(-i \frac{S(m)}{\hbar} \right)
\right|^2,
\end{eqnarray} 
where the notation $U(a)$ indicates the physical implementation of the unitary transformation on the state defined by the initial physical property $a$. Thus $P(b|U(a))$ has a well-defined operational meaning which is completely independent of the theoretical formalism. We can now relate this experimental fact directly to the observations of non-classical correlations in quantum paradoxes by noting that the spectral decomposition of the unitary operator $\hat{U}$ uses the same projectors that are also used to describe measurement outcomes. The complex probabilities of $P(m,b|a)$ are therefore related to the dynamical structure of unitary transformations according to 
\begin{equation}
\label{eq:trans}
P(b|U(a)) = \frac{1}{P(b|a)}\left| 
\sum_m P(m,b|a) \exp\left(-i \frac{S(m)}{\hbar} \right)
\right|^2.
\end{equation} 
This equation expresses a universally valid relation between the statistical correlations of non-commuting observables and the dynamics generated by the observables in question. Specifically, Eq.(\ref{eq:trans}) states that the statistical correlations between dynamically connected physical properties must correspond to complex and non-positive probabilities. It is thus possible to explain quantum paradoxes as a necessary consequence of the correct relation between non-commuting (and hence dynamically connected) physical properties given by Eq.(\ref{eq:trans}). Complex probabilities emerge as a result of the experimentally observable transformation dynamics between $a$ and $b$ generated by $m$. The emergence of non-positive probabilities is therefore not an arbitrary assumption, but instead represents a characteristic signature of the dynamical relations between the physical properties \cite{Hof11,Hof14a}. To achieve a more intuitive understanding, the analogy with classical phase space concepts may be helpful: $(a,m)$ and $(m,b)$ correspond to different phase space points separated by a distance along $m$ that can be covered by a transformation generated by $M$. This distance between $a$ and $b$ along $m$ can be expressed by an action $S_{\mathrm{opt.}}(m)$ that would maximize the probability $P(b|U(a))$ of getting from $a$ to $b$ by transformations along $m$. It can be seen from Eq.(\ref{eq:trans}) that the maximal statistical overlap is achieved whenever the action $S_{\mathrm{opt.}}(m)$ exactly compensates the original phase of the complex joint probability, 
\begin{equation}
\label{eq:opt}
S_{\mathrm{opt.}}(m,a,b) = \hbar \mbox{Arg}\left(P(m,b|a)\right). 
\end{equation}
The complex phase of $P(m,b|a)$ therefore describes the distance of transformation between $a$ and $b$ along $m$ \cite{Hof11}, where the gradient of the action $S_{\mathrm{opt.}}(m)$ in $m$ (that is, the difference between neighboring $m$) corresponds to the distance between $(a,m)$ and $(m,b)$ in the classical limit. It is therefore possible to understand the complex phases of the joint probabilities intuitively as an expression of the optimized transformation from $a$ to $b$ along $m$. Importantly, this is the only correct relation between the three physical properties in question, and should therefore replace the classical notion of determinism, where the value of the third property would given by an analytical function of the other two properties \cite{Hof12,Hof14a}.

The action itself can usually be expressed by a product of a generator with a measure of transformation distance: energy times time, angular momentum times angle, momentum times distance, and so on. In the context of quantum paradoxes, it is often convenient to consider spin systems, where the angular momentum around the z-axis is given by $L_z=\hbar m_z$ and the action of a rotation by an angle of $\phi$ around the $z$-axis is given by $S(m_z)=  \hbar m_z \phi$. If $b$ is obtained from $a$ by a rotation of angle $\phi$ around the $z$-axis, the action phases of the complex joint probabilities $P(m_z,b|a)$ are given by $S_{\mathrm{opt.}}= \hbar m_z \phi + S_{\mathrm{Norm}}$, where the correct value of $S_{\mathrm{Norm}}$ can be derived from the requirement that $P(b|a)=\sum_{m_z} P(m_z,b|a)$ is a positive number. Note that this requirement eliminates the artificial need for the definition of unphysical phases in the Hilbert space formalism and can even be used to explain the physics of gauge transformations, as previously explained in \cite{Hof14a}.

Although the imaginary parts of complex probabilities are only seen in the dynamics of transformations, the real parts appear as well-defined contributions in marginal distributions and in uncertainty limited joint measurements. In particular, the joint probabilities $P(m,b|a)$ fully define the marginal distributions $P(m|a)$ and $P(b|a)$ observed in separate measurements of $m$ and $b$. Quantum paradoxes mainly identify cases where the consequences of negative contributions in $P(m,b|a)$ are visible in the positive-values marginal distributions. Eq.(\ref{eq:trans}) shows that the contributions to the statistics of $a$, $b$ and $m$ will be negative whenever the action phases associated with the specific contribution are larger than $\pi/2$ \cite{Hof11}. In particular, action phases of zero and $\pi$ will result in a probability distribution with negative and positive real values. Such action phases describe a transformation that returns to its origin when it is applied twice, so that the transformation itself is half-periodic. We can therefore conclude that quantum mechanics predicts negative joint probabilities for the physical properties $a$, $b$ and $m$ if $a$ can be approximately transformed into $b$ by a half-periodic transformation generated by $m$. 

Quantum paradoxes are an immediate result of the negative statistical weights associated with half-periodic transformations according to the fundamental relations between the physical properties expressed by the Hilbert space formalism. It is therefore possible to explain quantum paradoxes by replacing the assumption of joint realities and positive probabilities with the correct relation between potential realities given by Eq.(\ref{eq:trans}). According to this relation, joint probabilities are {\it expected} to be non-positive, and the origin of negative probabilities can be traced directly to transformations between the physical properties involved in the paradox. In the following, I will apply this analysis to the various quantum paradoxes. In each case, it can be shown that the negative probabilities that result in the paradox originate from the action of half-periodic transformations that describe the relation between the physical properties involved in the paradox. Quantum paradoxes can therefore be explained by the transformation dynamics that defines the fundamental relations between the different properties of a physical system. 

\section{Negative probabilities in violations of Leggett-Garg inequalities}

\begin{figure}[th]
\begin{picture}(440,280)
\put(80,0){\makebox(280,280){\vspace*{-6cm}\hspace*{7.5cm}
\scalebox{1.2}[1.2]{
\includegraphics{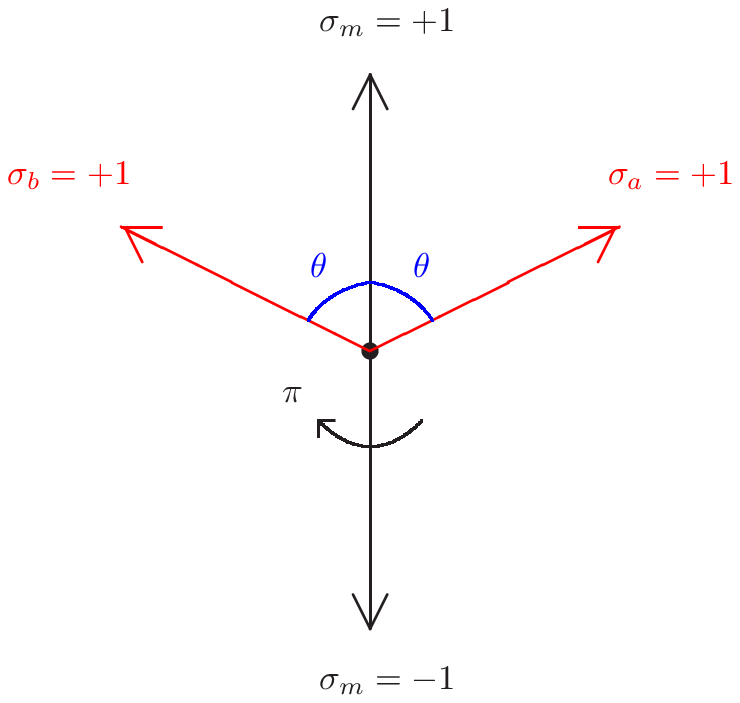}}}}
\end{picture}
\caption{\label{fig1} Illustration of the relation between the components of a spin-1/2 system that results in the violation of the Leggett-Garg inequality. The negative probability of $\sigma_m=-1$ is a direct consequence of the half-periodic spin-flip around the $m$-axis.}
\end{figure}

Perhaps the most direct connection between joint probabilities and inequality violations is given by the Leggett-Garg inequalities. Originally, these inequalities were formulated as a temporal equivalent to Bell's inequalities. However, the dynamics of a two-level system always corresponds to a spin precession, so correlations of the same observable at different times correspond to correlations between different spin directions at the same time. Therefore, Leggett-Garg inequalities essentially describe the limits that realism imposes on the values of three different spin directions of a two-level system, given by $\sigma_a=\pm 1$, $\sigma_m=\pm 1$ and $\sigma_b=\pm 1$ \cite{LGI}. Fig. \ref{fig1} illustrates this relation between three different spin directions for the symmetric case of equal angles between $\sigma_a$ and $\sigma_m$ and between $\sigma_m$ and $\sigma_a$.

If a specific value of $\sigma_a$ is chosen as initial condition, the joint probabilities of $\sigma_m$ and $\sigma_b$ can be determined from the operator algebra. Specifically, it is possible to derive joint probabilities of $\sigma_m$ and $\sigma_b$ by using the expectation values of $\sigma_m$, $\sigma_b$ and their operator product, e.g.
\begin{equation}
\label{eq:LGI}
P(\sigma_m=-1, \sigma_b=+1|\sigma_a=+1) = \frac{1}{4}\left( 1+\langle \sigma_b \rangle_{\sigma_a} - \langle \sigma_m \rangle_{\sigma_a} - \langle \sigma_m \sigma_b \rangle_{\sigma_a}\right). 
\end{equation}
In a two level system, the expectation value of $\sigma_m$ for an initial spin of $\sigma_a=+1$ is given by $\cos (\theta)$, where $\theta$ is the angle between the directions of the spins. If the three spins are equally spaced and lie in the same plane, so that the angle between $\sigma_a$ and $\sigma_b$ is $2 \theta$ and $\sigma_m$ is exactly in the middle between $\sigma_a$ and $\sigma_b$, the joint probability predicted from separate measurements of the expectation values is 
\begin{eqnarray}
\label{eq:negprob}
P(\sigma_m=-1, \sigma_b=+1|\sigma_a=+1) &=& \frac{1}{4}(1 + \cos(2\theta) - \cos(\theta)-\cos(\theta))
\nonumber \\ &=& \frac{1}{2} \cos(\theta) \; (\cos(\theta)-1).
\end{eqnarray}
This probability is negative for all values of $\theta<\pi/2$, so that the Leggett-Garg inequality for the correlations in Eq.(\ref{eq:LGI}) is violated whenever the positive spins are in the upper half of the Bloch sphere. On the other hand, the specific value of the negative probability is a function of the angle between the spins, with a maximal violation of Legett-Garg inequalities at $\cos(\theta)=1/2$, where the joint probability is $-1/8$. 

As explained in the previous section, the appearance of negative joint probabilities can be traced back to half-periodic transformations between the physical properties in question. As indicated in Fig. \ref{fig1}, $\sigma_a=+1$ can be transformed into $\sigma_b=+1$ by a 180 degree rotation around the axis given by $\sigma_m$.
Since the action of the spin-flip around the $m$-axis can be given by $S_{+1}=0$ and $S_{-1}=\pi \hbar$, the sum over $m$ in Eq.(\ref{eq:trans}) is equal to the difference between the two joint probabilities for $(\sigma_m=+1,\sigma_b=+1)$ and $(\sigma_m=-1,\sigma_b=+1)$. Since the overlap after the transformation is $P(\sigma_b=+1|U(\sigma_a=+1))=1$, the difference between the joint probabilities is
\begin{equation}
\label{eq:diff}
P(\sigma_m=+1, \sigma_b=+1|\sigma_a=+1) - P(\sigma_m=-1, \sigma_b=+1|\sigma_a=+1)
= \sqrt{P(\sigma_b=+1|\sigma_a=+1)}. 
\end{equation}
By definition, the sum of the joint probabilities is given by the marginal probability $P(\sigma_b=+1|\sigma_a=+1)$. The joint probability of $\sigma_m=-1$ and $\sigma_b=+1$ can therefore be derived from the directly observable probability $P(\sigma_b=+1|\sigma_a=+1)$ by
\begin{equation}
\label{eq:result}
P(\sigma_m=-1, \sigma_b=+1|\sigma_a=+1) = 
\frac{1}{2} \left(P(\sigma_b=+1|\sigma_a=+1) - \sqrt{P(\sigma_b=+1|\sigma_a=+1)}\right). 
\end{equation}
Since $P(\sigma_b=+1|\sigma_a=+1)=\cos^2(\theta)$, this is equal to the result derived from the expectation values in Eq.(\ref{eq:negprob}). 

The essential advantage of the derivation from Eq.(\ref{eq:trans}) is that it  derives the negative probability from the familiar physics of rotations around the $m$-axis. In close analogy to classical physics, this action is basically given by the product of angular momentum and angle, with an additional offset that is needed to ensure that the marginal probabilities are positive and real. However, the laws of quantum physics require that this action also appears as a phase in the complex probabilities that describe the statistics of the spins, and it is this phase difference of $\pi$ that explains the negative probabilities responsible for Leggett-Garg inequality violations.

\section{Three box paradox and quantum Cheshire cats}

In the Leggett-Garg scenario, the limitation to a single two-level system means that we can easily understand the physics of half-periodic transformations. In multi-level systems, there is a much wider range of half-periodic transformations, so there is a much wider variety of quantum paradoxes that can be constructed from the corresponding negative probabilities. A particularly, illustrative example is the three box paradox, where the initial state is defined in terms of a superposition of three possible paths represented by numbered boxes \cite{box},
\begin{equation}
\mid a \rangle =\frac{1}{\sqrt{3}} \left(\mid 1 \rangle + \mid 2 \rangle + \mid 3 \rangle \right).
\end{equation}
After the system passes through the three boxes, it is measured in a different superposition given by 
\begin{equation}
\mid b \rangle =\frac{1}{\sqrt{3}} \left(\mid 1 \rangle + \mid 2 \rangle - \mid 3 \rangle \right).
\end{equation}
The paradox is based on the observation that $\mid 2 \rangle - \mid 3 \rangle$ is orthogonal to $\mid a \rangle$ and $\mid 2 \rangle + \mid 3 \rangle$ is orthogonal to $\mid b \rangle$. If we assume that we can independently eliminate possibilities that contradict either $a$ or $b$, it seems that only box 1 is left as a possible path between $a$ and $b$. However, the same argument can be made to exclude box 1 and box 3, leaving only box 2 as a possible path between $a$ and $b$. 

Here, the resolution of the paradox by Eq.(\ref{eq:trans}) is particularly direct, since the formulation of the problem defines the relation between $a$ and $b$ in terms of a half-periodic transformation with actions of $S(1)=S(2)=0$ and $S(3)=\hbar \pi$. The marginal probability $P(b|a)=1/9$ requires that
\begin{equation}
P(1,b|a) + P(2,b|a) + P(3,b|a) = \frac{1}{9}.
\end{equation}
In addition, Eq.(\ref{eq:trans}) requires that 
\begin{equation}
9 \left| P(1,b|a) + P(2,b|a) - P(3,b|a) \right|^2 = 1.
\end{equation}
Since box 1 and box 2 are symmetric, we also know that $P(1,b|a)=P(2,b|a)$, so the joint probabilities are given by
\begin{equation}
P(1,b|a) = \frac{1}{9},
\hspace{0.5cm}
P(2,b|a) = \frac{1}{9},
\hspace{0.5cm}
P(3,b|a) = - \frac{1}{9}.
\end{equation}
The probability of zero for the superpositions of box 2 and box 3 can be confirmed, but it originates from a cancellation of positive and negative joint probabilities. The assignment of paths fails because the relation between the physical properties $a$, $b$ and the boxes is given by the transformation between $a$ and $b$ generated by applying a box-dependent action. 

The logic of the three box paradox nicely illustrates the role of half-periodic transformations, since the initial and the final state are defined only in terms of the phase relations of their components in the intermediate basis. In fact, it is possible to construct a large number of quantum paradoxes in this manner. If a single photon travels through a multi-path interferometer from an input port $a$ to an output port $b$ such that the initial probability of finding the photon in $b$ is 1, phase shifts of $\pi$ induced in a specific selection of intermediate paths will not only reduce the output probability in $b$, but actually result in negative joint probabilities for the intermediate paths $m$ and the output $b$.

\begin{figure}[th]
\begin{picture}(440,180)
\put(0,0){\makebox(440,180){\vspace*{-6cm}\hspace*{3.5cm}
\scalebox{0.8}[0.8]{
\includegraphics{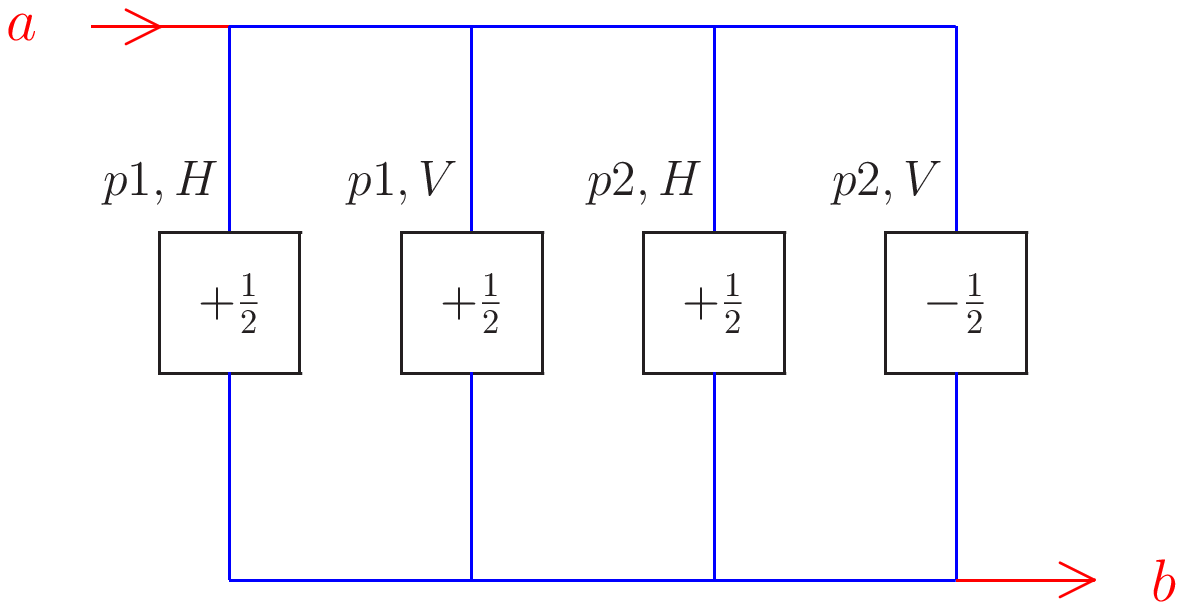}}}}
\end{picture}
\caption{\label{fig2} Explanation of the Cheshire cat paradox as an extension of the three box paradox to four boxes. The transformation from $a$ to $b$ is achieved by an action of $S(p2,V)=\pi \hbar$. The ``cat'' appears to be in $p1$, but the ``smile'' given by the probability difference $P(H)-P(V)$ of the polarizations is found in $p2$.}
\end{figure}

An example of a quantum paradox that can be constructed in this manner is the Cheshire cat scenario, where the polarization of a photon appears to be separated from the spatial path that the photon took \cite{Che,Den14}. The paradox is easy to explain once the complete statistics of polarization and spatial paths is considered. Fig. \ref{fig2} illustrates the relation with the three box paradox by assigning a separate ``box'' to each combination of path and polarization inside the interferometer. The photon can be found in the spatial paths $p1$ or $p2$ with a horizontal polarization $H$ or a vertical polarization $V$. If the initial state is an equal superposition of all four possibilities, and the final state is generated by a phase shift of $\pi$ in the $V$-polarized component of path $p2$ followed by a detection of an equal superposition in the output, the joint probabilities of paths and polarizations inside the interferometer are given by 
\begin{eqnarray}
P(p1,H;b|a) = \frac{1}{8},
\hspace{0.5cm} &&
P(p1,V;b|a) = \frac{1}{8},
\nonumber \\
P(p2,H;b|a) = \frac{1}{8},
\hspace{0.5cm} &&
P(p2,V;b|a) = - \frac{1}{8}.
\end{eqnarray}
It is tempting to consider the polarization and the paths separately, in which case it seems that the probability of $p1$ is one and that of $p2$ is zero, while the probability of $H$-polarization is $1$ and that of $V$ polarization is zero. However, the correlation between polarization and path then seems to indicate that the $H$-polarization propagates only along $p2$, even though the photon only propagates along path $p1$. 

Importantly, there is no direct measurement of path or polarization, since the initial and final states are not eigenstates of the corresponding properties. Similar to the three box paradox, information about the path and the polarization is induced based on the relations of the intermediate eigenstates with either the initial state or the final state, one of which now expresses a particular correlation between path and probability. To resolve the paradox, it is important to realize that this induction is based on false assumptions about the relations between the initial and final conditions $a$, $b$ and the intermediate properties of path and polarization.

\section{The Hardy paradox}

Like the Cheshire cat paradox, the Hardy paradox is based on two-path interferometers \cite{Har92}. Two particles are sent through two different interferometers set up in parallel, with outer paths $O$ and inner paths $I$. However, the inner paths cross each other, so that an interaction effectively eliminates the particles if they are both in the inner paths. As a result of this interaction, the interference between the two paths is disturbed in both interferometers, and the output port can switch from the original output port to the opposite port. 

The situation after the interaction has eliminated the combination $(I_1,I_2)$ can be described by an equal superposition of the three remaining combinations of paths,
\begin{equation}
\mid a \rangle = \frac{1}{\sqrt{3}}\left(\mid O_1,O_2 \rangle + \mid O_1,I_2 \rangle + \mid I_1,O_2 \rangle\right). 
\end{equation}
The detection of a photon in the port opposite to the original output port is described by a negative superposition of the paths,
\begin{equation}
\mid b_i \rangle = \frac{1}{\sqrt{2}}\left(\mid O_i \rangle - \mid I_i \rangle\right). 
\end{equation}
The probability of finding both particles in the opposite output ports is $P(b_1,b_2|a)=1/12$. However, it seems that a switch to the opposite port requires the presence of the other particle in the inner path, as shown by the directly observable probabilities $P(b_1,O_2|a)=0$ and $P(O_1,b_2|a)=0$. Assuming a measurement independent joint reality of paths and output ports, one would think that the outcome $(b_1,b_2)$ is only possible if both particles took the inner path, $(I_1,I_2)$. However, this option was clearly eliminated in the preparation of $a$. 

The probability $P(b_1,b_2|a)$ can be expressed as a sum of joint probabilities,
\begin{eqnarray}
\label{eq:psum}
P(b_1,b_2) &=& P(O_1,O_2;b_1,b_2|a) + P(O_1,I_2;b_1,b_2|a) 
\nonumber \\ &&
+P(I_1,O_2;b_1,b_2|a) + P(I_1,I_2;b_1,b_2|a).
\end{eqnarray}
Because of the preparation condition, we know that $P(I_1,I_2;b_1,b_2|a)=0$. In addition, we can impose the condition that $P(b_1,O_2|a)=0$ and $P(O_1,b_2|a)=0$ by requiring that the corresponding contributions of the joint probabilities sum up to zero,
\begin{eqnarray}
\label{eq:zero}
P(O_1,O_2;b_1,b_2|a) + P(O_1,I_2;b_1,b_2|a) = 0, 
\nonumber \\
P(O_1,O_2;b_1,b_2|a) + P(I_1,O_2;b_1,b_2|a) = 0.
\end{eqnarray}
With these conditions, it is possible to derive the values of the joint probabilities in the Hardy paradox,
\begin{eqnarray}
P(O_1,O_2;b_1,b_2|a)= -\frac{1}{12}, \hspace{0.5cm}&&
P(O_1,I_2;b_1,b_2|a)= \frac{1}{12} ,
\nonumber \\
P(I_1,O_2;b_1,b_2|a)= \frac{1}{12}, \hspace{0.5cm}&&
P(I_1,I_2;b_1,b_2|a)= 0.
\end{eqnarray}
Thus the negative probability of $(O_1,O_2)$ ensures that the individual probabilities of $(O_1)$ and $(O_2)$ can be zero even though $(O_1,I_2)$ and $(I_1,O_2)$ contribute positively to the probability of the observed output port combination $(b_1,b_2)$. 

As in the previous examples, the negative probability in the Hardy paradox can be explained in terms of the transformation dynamics in the paths $m$ that relates the initial state $a$ to the final state $b$. According to Eq.(\ref{eq:trans}),
\begin{equation}
P(O_1,O_2;b_1,b_2|a) - P(O_1,I_2;b_1,b_2|a) - P(I_1,O_2;b_1,b_2|a)
 = - \sqrt{P(b_1,b_s|a) P(b_1,b_2|U(a))},
\end{equation}
where the maximal overlap of $P(b|U(a))=3/4$ is achieved by applying a phase shift of $\pi$ between the arms of each interferometer. Together with the condition given by Eq.(\ref{eq:psum}), this relation determines the negative value of $P(O_1,O_2;b_1,b_2|a)$ that is responsible for the paradoxical statistics observed in separate measurements of paths and outputs. Thus the Hardy paradox can also be traced back to a half-periodic transformation that relates the initial conditions $a$ with the final conditions $b$, represented in this case by a phase flip in both of the two interferometers. 

\section{Correlations and contextuality}

An interesting class of quantum paradoxes is associated with the concept of contextuality, which is a formalized version of the notion that different measurements do not refer to the same reality. In fact, it has recently been suggested that the ``strange'' results of weak measurements should be understood as evidence of contextuality \cite{Pus14}, which seems to be similar to the observation that the fundamental relations between physical properties described by Eq.(\ref{eq:trans}) indicate that these results cannot have a joint reality. A particularly interesting illustration of the concept can be obtained by considering the orthogonal components of two spin-1/2 systems, where the components $X_i$, $Y_i$, $Z_i$ of spin $i$ have values of $\pm 1$ \cite{KS,Mer90}. Each of the spin components of spin 1 commutes with each of the components of spin 2. On the other hand, the products $X_1 X_2$, $Y_1 Y_2$ and $Z_1 Z_2$ also commute with each other, as do the products $X_1 Y_2$, $Y_1 X_2$ and $Z_1 Z_2$. For two local spin measurements, the value of the product is obtained by multiplying the individual results, e.g.
\begin{equation}
\label{eq:xcorr}
(X_1 X_2) = (X_1) (X_2)
\end{equation}
If we assume that the values obtained from measurements of ($X_1$, $X_2$) and from ($Y_1$,$Y_2$) are the same as those obtained from measurements of ($X_1$, $Y_2$) and ($X_1$,$X_2$), we should conclude that the products of the values are also identical,
\begin{equation}
\label{eq:prodcorr}
(X_1 X_2)(Y_1 Y_2) = (X_1 Y_2)(Y_1 X_2).
\end{equation}
Of course it is not actually possible to obtain the outcomes of the measurement at the same time, so it is not possible to check the relation in Eq.(\ref{eq:prodcorr}) by a direct measurement. However, it is possible to measure $X_1 X_2$, $Y_1 Y_2$ and $Z_1 Z_2$, since these three products commute with each other. If the three correlations were independent of each other, there should be eight possible combinations of eigenvalues. However, there are only four, and all four of them satisfy the relation
\begin{equation}
\label{eq:corr1}
(X_1 X_2) (Y_1 Y_2) = - (Z_1 Z_2).
\end{equation}
It thus seems as if the product $Z_1 Z_2$ is completely determined by the product of the local $X_i$ and $Y_i$. However, the situation changes drastically when the products $X_1 Y_2$, $Y_1 X_2$ and $Z_1 Z_2$ are measured instead. Simply by re-arranging the distribution of the $X_i$ and $Y_i$, the sign of the product changes to 
\begin{equation}
\label{eq:corr2}
(X_1 Y_2) (Y_1 X_2) = (Z_1 Z_2),
\end{equation}
violating Eq.(\ref{eq:prodcorr}) and hence the context independence of the products of local spins. Any measurement independent assignment of the four local spin components $X_i$ and $Y_i$ can only satisfy either Eq.(\ref{eq:corr1}) or Eq.(\ref{eq:corr2}), never both. Thus the experimentally observed correlation products of spin systems provide a particularly striking example of a contextuality paradox.

We can now analyze the contextuality paradox of a pair of spins by considering the actual relations between the non-commuting physical properties according to Eq.(\ref{eq:trans}). Importantly, the measurement context plays a fundamental role in Eq.(\ref{eq:trans}), since the dynamics of transformations between $a=(X_1,X_2)$ and $b=(Y_1,Y_2)$ is different from the dynamics of transformations between $a=(X_1,Y_2)$ and $b=(Y_1,X_2)$. Oppositely, the precise choice of values is not important, since the relations given by Eq.(\ref{eq:corr1}) and Eq.(\ref{eq:corr2}) apply equally to any combination of values. The contextuality paradox arises because the assumption of a joint reality is used to calculate the products $X_1 X_2$ and $Y_1 Y_2$ from the values assigned to $a=(X_1,Y_2)$ and $b=(Y_1,X_2)$, even though there is no joint reality to the properties $a$, $b$ and $m=(X_1 X_2, Y_1 Y_2)$. While it is technically correct to say that $m$ belongs to a context different from $a$ and $b$, the essential point is that the physical property $m$ is the generator of a half-periodic transformation that transforms $a=(X_1,Y_2)$ into $b=(Y_1,X_2)$. As a result, the products $X_1 X_2$ and $Y_1 Y_2$ are related to $a$ and $b$ by complex probabilities, and not by products of local values taken separately from $a$ and from $b$. 

To illustrate how the dynamics generated by $m$ explains the contextuality paradox, it is useful to derive the half-periodic transformation for a specific combination of $a$ and $b$. For simplicity, I consider the transformation between the initial condition $a=(X_1=+1,Y_2=+1)$ and the final condition $b=(Y_1=+1,X_2=+1)$. These two conditions are related by a swap of system 1 and system 2. Since a double swap restores the original situation, the swap operation is a half-periodic transformation. When written as a unitary transformation, the eigenstates of the swap operations are the anti-symmetric state $S$ characterized by $m = (X_1 X_2 =-1, Y_1 Y_2=-1)$, and the symmetric states $T_x$, $T_y$, $T_z$, characterized by $m = (X_1 X_2 =-1, Y_1 Y_2=+1)$ for $T_x$, $m = (X_1 X_2 =+1, Y_1 Y_2=-1)$ for $T_y$, and $m = (X_1 X_2 =+1, Y_1 Y_2=+1)$ for $T_z$. The swap operation assigns a phase of $\pi$ to the anti-symmetric state $S$, so the joint probabilities derived from Eq.(\ref{eq:trans}) are
\begin{eqnarray}
\label{eq:context}
P(S;b|a)=-\frac{1}{8}, \hspace{0.5cm} &&
P(T_x;b|a)= \frac{1}{8}, 
\nonumber \\ 
P(T_y;b|a)= \frac{1}{8}, \hspace{0.5cm} &&
P(T_z;b|a)= \frac{1}{8}.
\end{eqnarray}
It is now possible to explain the contradiction between Eq.(\ref{eq:corr1}) and Eq.(\ref{eq:corr2}) and the contextuality paradox associated with it. Specifically, Eq.(\ref{eq:corr1}) refers to the correlations that define the physical property $m=S,T_x,T_y,T_z$, while Eq.(\ref{eq:corr2}) refers to the correlations between $a$ and $b$. Even though the relation between $a$ and $b$ suggests well-defined values of $X_1 X_2 =+1$, $Y_1 Y_2=+1$, and (according to Eq.(\ref{eq:corr2})) $Z_1 Z_2 =+1$, none of the four $m$ fits all of these conditions. Instead, $S$ has opposite values for all three products, while $T_x$, $T_y$ and $T_z$ have opposite values for one of the three each. The expected relation between $X_1$ from $a$, $X_2$ from $b$, and $X_1 X_2$ from $m$ only holds for the average of $X_1 X_2$, which is $+1$ because the probabilities of $S$ and $T_x$ sum up to zero. In general, Eq.(\ref{eq:corr2}) applies to the conditional averages of the individual products $X_1 X_2$, $Y_1 Y_2$ and $Z_1 Z_2$ obtained from the joint probabilities $P(m;b|a)$.

Importantly, it is not correct to derive the value of $(X_1 X_2)$ from the results of two separate non-commuting measurements that happen to include the value of $X_1$ in measurement $a$ and the value of $X_2$ in measurement $b$. The correct relation between three different measurements is determined by the action of transformations between them, and the classical product relation does not apply when $X_1$, $X_2$, and $(X_1 X_2)$ belong to three different measurements that cannot be performed jointly. Thus, Eq.(\ref{eq:trans}) gives a more explicit meaning to the notion of contextuality in quantum mechanics.

\section{Violation of Bell's inequalities by contextual correlations}

The insights gained about the contextuality of correlations can also be used to shed some light on what is arguably the best know quantum paradox, the violation of Bell's inequalities \cite{Bell}. Essentially, Bell's inequalities describe the limit of correlations between the local spin components $X_i$ and $Y_i$ when each component has a value of $\pm 1$. The limit is given by a sum of four individual correlations,
\begin{equation}
\label{eq:K}
K= X_1 X_2 + X_1 Y_2 + Y_1 X_2 - Y_1 Y_2.
\end{equation}
It is easy to see that, for any combination of spin values, the value of $K$ is either $+2$ or $-2$. Therefore, all positive valued joint probability distributions for $X_i$ and $Y_i$ satisfy the Bell's inequality
\begin{equation}
|\langle K \rangle | \leq 2.
\end{equation}
Oppositely, a violation of Bell's inequality corresponds to the assignment of a negative joint probability to $K=-2$ for $\langle K \rangle>2$, or to $K=+2$ for $\langle K \rangle<-2$.

Due to the contextuality of spin correlations there are two non-equivalent ways to define joint probabilities for pairs of spins, depending on the combinations of spins in the conditions $m$ and $b$. We can either chose $m=(X_1,X_2)$ and $b=(Y_1,Y_2)$, or we can chose $m^\prime=(X_1,Y_2)$ and $b^\prime=(Y_1,X_2)$. In the former case, the correlations $(X_1 X_2)$ and $(Y_1 Y_2)$ are directly defined by $m$ and $b$, respectively, but the correlations $(X_1 Y_2)$ and $(Y_1 X_2)$ will depend on the transformation dynamics of $b$ generated by $m$. In the latter case, the transformation dynamics will show up in $(X_1 X_2)$ and $(Y_1 Y_2)$, while $(X_1 Y_2)$ and $(Y_1 X_2)$ are directly defined by $m$ and $b$. 

In general, negative probabilities occur if there is a half-periodic transformation that transforms $b$ into $a$ along $m$. To explain the violation of Bell's inequalities, we need a half-periodic transformation that transforms product states into entangled states, where the eigenstates themselves should also be product states. A well-known half-periodic transformation with these properties is the quantum-controlled NOT, which essentially describes conditional spin-flips in the two systems. Specifically, a quantum-controlled NOT with eigenstates $m=(X_1,X_2)$ can transform eigenstates of $b=(Y_1,Y_2)$ into entangled states with maximal correlations of $X_1 Y_2=\pm1$ and $Y_1 X_2=\pm 1$. We can therefore express the initial state $a(0)$ with well-defined correlations of $X_1 Y_2=+1$ and $Y_1 X_2=+1$ in terms of its non-positive joint probabilities of $m=(X_1,X_2)$ and $b=(Y_1,Y_2)$ according to the transformation rules of Eq.(\ref{eq:trans}),
\begin{eqnarray}
\label{eq:azero}
P(m=(+1,+1);b=(+1,+1)|a(0))&=&\frac{1}{8}, 
\nonumber \\
P(m=(+1,-1);b=(+1,+1)|a(0))&=& \frac{1}{8}, 
\nonumber \\
P(m=(-1,+1);b=(+1,+1)|a(0))&=& \frac{1}{8}, 
\nonumber \\
P(m=(-1,-1);b=(+1,+1)|a(0))&=& -\frac{1}{8}.
\end{eqnarray}
Comparable sets of probabilities can be obtained for all other values of $b=(Y_1,Y_2)$. Although the joint probabilities already include negative values, this state does not violate Bell's inequalities, since the negative and the positive contributions cancel out in the expectation values of $(X_1 X_2)$ and $(Y_1 Y_2)$, resulting in expectation values of zero for these correlations. The negative probabilities obtained from the half-periodic transformations between $a(0)$ and $b$ along $m$ are not sufficient to achieve a violation of Bell's inequalities all by themselves. However, Eq. (\ref{eq:trans}) does not require a perfect transformation with $p(b|U(a))=1$. Instead, it is sufficient if the half-periodic transformation optimizes the overlap between the initial condition $a$ and the final condition $b$. It is therefore possible to increase the correlation $(X_1 X_2 - Y_1 Y_2)$ in $a$ without eliminating the negative probabilities defined by the transformation of $(Y_1,Y_2)$ into $(X_1 Y_2,Y_1 X_2)$.

The optimization can be achieved by choosing an initial condition $a(\theta)$ defined by correlations between spin-directions in the $XY$-plane with an angle of $\theta$ between them,
\begin{equation}
\label{eq:init}
\cos (\theta) X_1 Y_2 + \sin (\theta) X_1 X_2 = 1, \hspace{0.5cm}
\cos (\theta) Y_1 X_2 - \sin (\theta) Y_1 Y_2 = 1.
\end{equation}
The complex joint probabilities defined by Eq.(\ref{eq:trans}) now include both the negative probabilities of the conditional spin flips and the increased probabilities for $X_1 X_2 = +1$ in $m$ and for $Y_1 Y_2 = -1$ in $b$. For $b=(+1,+1)$, the joint probabilities then read
\begin{eqnarray}
\label{eq:atheta}
P(m=(+1,+1);b=(+1,+1)|a(0))&=&\frac{1}{8} \cos(\theta), 
\nonumber \\
P(m=(+1,-1);b=(+1,+1)|a(0))&=& \frac{1}{8} (1-\sin(\theta)), 
\nonumber \\
P(m=(-1,+1);b=(+1,+1)|a(0))&=& \frac{1}{8} (1-\sin(\theta)), 
\nonumber \\
P(m=(-1,-1);b=(+1,+1)|a(0))&=& -\frac{1}{8} \cos(\theta).
\end{eqnarray}
Among these probabilities, only $m=(+1,+1)$ contributes to $K=+2$. Therefore, the negativity of $P(K=-2)$ can be increased by reducing the positive probabilities of $m=(+1,-1)$ and $m=(-1,+1)$. The maximal Bell's inequality violation is obtained by finding the maximal negative value of the sum of all probabilities for outcomes with $K=-2$. 

\begin{table}[h]
\begin{tabular}{|cc|cccc|}
\hline
&&\multicolumn{4}{c|}{$b=(Y_1,Y_2)$}\\
$m=(X_1,X_2)$ &&$(-1,-1)$&$(+1,-1)$&$(-1,+1)$&$(+1,+1)$\\[0.1cm]
\hline
$(-1,-1)$&& \hspace*{0.3cm}$\frac{1}{8} \cos (\theta)$ & 
            $\frac{1}{8} (1+\sin (\theta))$&
            $\frac{1}{8} (1+\sin (\theta))$& 
            $- \frac{1}{8} \cos (\theta)$\\
$(+1,-1)$&& $\frac{1}{8} (1-\sin (\theta))$ & $-\frac{1}{8} \cos (\theta)$
     &\hspace*{0.3cm}$\frac{1}{8} \cos (\theta)$ & $\frac{1}{8} (1-\sin (\theta))$\\
$(-1,+1)$&& $\frac{1}{8} (1-\sin (\theta))$ & \hspace*{0.3cm}$\frac{1}{8} \cos (\theta)$
     &$-\frac{1}{8} \cos (\theta)$ & $\frac{1}{8} (1-\sin (\theta))$\\
$(+1,+1)$&& $- \frac{1}{8} \cos (\theta)$ & $\frac{1}{8} (1+\sin (\theta))$
     &$\frac{1}{8} (1+\sin (\theta))$ & \hspace*{0.3cm}$\frac{1}{8} \cos (\theta)$\\[0.1cm]
\hline
\end{tabular}

\caption{ \label{table1} Joint probabilities for the local spins under the initial conditions given in Eq.(\ref{eq:init}). Correlations between $X_1$ and $Y_2$, and between $Y_1$ and $X_2$ correspond to specific combinations of lines and columns. Negative probabilities appear for all combinations with $X_1Y_2=-1$ and $Y_2 X_2=-1$.}
\end{table}

The complete set of joint probabilities is shown in table \ref{table1}. It might be worth noting that the $\theta=\pi/4$ case of these non-positive probabilities has recently been observed experimentally using weak measurements \cite{Hig15}, which confirms the basic consistency between quantum paradoxes and the joint statistics given by Eq. (\ref{eq:jointprob}). Probabilities of $(1+\sin(\theta))/8$ are assigned to all combinations with $X_1 X_2=+1$ and $Y_1 Y_2=-1$. For all of these combinations, $K=+2$. Likewise, probabilities of $(1-\sin(\theta))/8$ are assigned to combinations with $X_1 X_2=-1$ and $Y_1 Y_2=+1$, which have $K=-2$. For the remaining eight combinations, positive probabilities of $\cos(\theta)/8$ are assigned to combinations with $X_1 Y_2=+1$ and $Y_1 X_2=+1$, where $K=+2$, and negative probabilities of $-\cos(\theta)/8$ are assigned to combinations with $X_1 Y_2=-1$ and $Y_1 X_2=-1$, where $K=-2$. In summary, probabilities are low but positive if the value of $K=-2$ can be obtained directly from $m=(X_1,X_2)$ and $B=(Y_1,Y_2)$, while the probabilities are negative if the value of $K=-2$ originates from correlations between $m$ and $b$. Bell's inequality is violated because the total probabilities corresponding to correlation sums of $K=-2$ have a negative value of
\begin{equation}
P(K=-2) = \frac{1}{2}\left(1-\sin(\theta)-\cos(\theta)\right) <0. 
\end{equation}
Thus, the violation of Bell's inequality is achieved by a combination of two local contexts in the initial correlations $a$, where one context is given by the choice of $m$ and $b$ and the other context is given by the correlations between $m$ and $b$ that are determined by the dynamics of transformations. Negative probabilities appear in the correlations between $m$ and $b$ as a result of the half-periodic transformations that describe the relation between the initial condition $a$ and the properties $m$ and $b$, while the positive probabilities of contributions to $K=-2$ that are determined individually by $m$ and $b$ are sufficiently lower, resulting in an overall violation of the positive probability limit of $\langle K \rangle \leq 2$. 

\section{Why non-positive probabilities make sense}

To actually resolve quantum paradoxes, it is necessary to understand why the fundamental relations between physical properties should be expressed in terms of complex joint probabilities, where the complex phase is determined by the action of transformations between the different properties as given by Eq.(\ref{eq:trans}). Importantly, complex probabilities indicate that the fundamental relations between physical properties do not permit any assignment of joint reality to the physical properties. Obviously, this is not a trivial matter - we normally think of objects in terms of a complete set of physical properties, regardless of whether these properties are observed in a measurement or not. How can we explain this limitation of reality to properties that are actually measured?

It seems to be essential that complex probabilities relate to transformations that describe the actual dynamics generated by a physical property. We need to remind ourselves that physical reality is only known from interactions, and these interactions necessarily involve transformations of the type that define the phases of complex probabilities. In the classical limit, we can approximately separate the effect of the object from the changes of the object that are inadvertently caused by the interactions, because the changes caused by the scattering of light or by physical touch are sufficiently small to be neglected. The quantum formalism indicates that this approximation breaks down at the level of $\hbar$. It is therefore fundamentally impossible to separate the changes caused by the interaction process from the effects that define a physical object as a thing that we can see and touch. 

In direct measurements of a physical property, the interaction completely randomizes the unobserved properties, thereby eliminating the possibility of joint measurements. However, weak measurements can be used to obtain statistical evidence of the unobserved properties before the strong measurement randomized them. It is therefore possible to experimentally confirm the negative probabilities predicted by Eq.(\ref{eq:trans}), and corresponding experiments have already been reported for most of the paradoxes analyzed above \cite{Res04,Jor06,Tol07,Wil08,Lun09,Yok09,Gog11,Suz12}. The discussion presented here is intended as an explanation of the general physical principles that are revealed by these experimentally confirmed results. Both theory and experiment strongly indicate that there is no ambiguity in the relation between non-commuting physical properties. Paradoxes arise because we want to hold on to the fiction of a simultaneous reality even though there is no evidence in favour of it, and a lot of evidence against. It would seem to me that the realization that quantum physics integrates the structure of the dynamics into the statistics of the measurement outcomes is the key to a better understanding of the physics that results in this failure of fictional realism.  

Complex probabilities make sense because their complex phases have a well-defined operational meaning: they describe the relations between physical properties in terms of the dynamics of transformations between them. This relation replaces the assumption of a joint reality, which has no foundation in observable fact and fails to explain the statistics observed in quantum paradoxes. Importantly, Eq.(\ref{eq:trans}) is not an expression of randomness, but of a universal and fully deterministic relation between the three physical properties. It therefore replaces the causality relations of classical physics, which emerge only in the limit of low resolution \cite{Hof12}. These causality relations can explain the relation between different measurements in terms of negative joint probabilities, where negative values necessarily appear as a direct consequence of the transformations between them. Quantum paradoxes can therefore be resolved by properly understanding that the quantum formalism defines the relations between physical properties by the dynamics of their transformations, and not by a hypothetical joint reality that is never observed in any experiment.

\section{Conclusions}

In physics, all paradoxes are the consequence of wrong assumptions about the laws of nature. Once the physics is properly understood, paradoxes can be resolved by a correct explanation of the fundamental relations between the actual phenomena. Up to now, quantum paradoxes have remained mysterious because the formalism that actually predicts them was treated as a mathematical black box without any physical meaning. In this paper, I have shown that a better understanding of the physics is possible: Quantum paradoxes can all be explained by a single fundamental relation between the physical properties involved in the paradox, and this fundamental relation predicts negative joint probabilities whenever three physical properties are related to each other by half-periodic transformations. Importantly, this fundamental relation also explains why the expectation of a measurement independent reality fails: the reality of an object only emerges in interactions, so it is entirely possible that the interactions themselves are an inseparable part of this reality. By describing the relation between physical properties in terms of complex probabilities, quantum mechanics objectively identifies the correct relation between dynamics and reality valid for all phenomena at the quantum level \cite{Hof14a}. Quantum paradoxes can thus be resolved by a fundamental explanation of the actual physics described by the quantum formalism.

\section*{Acknowledgment}
This work was supported by JSPS KAKENHI Grant Number 24540427.

\end{document}